\documentstyle [epsf,12pt]{article}
\input epsf
\begin{document}
\hoffset -1 true cm
\textwidth 6.5 true in
\textheight 9.5 true in
\begin{center}
\title: {\bf Negative Energy Condition and Black Holes on the Brane}\\
\vspace{0.3cm}
\author:{ Naresh Dadhich}\\
 {IUCAA, Post Bag 4, Ganeshkhind, Pune 411007, India.}
\end{center}

\begin{abstract}
 We propose that for non-localizable energy distribution the relevant 
energy condition is determined by the gravitational field energy which is 
negative for positive non-gravitational energy. That is negativity of the 
non-localized energy is the "positive" energy condition. This would have 
direct application and relevance for a black hole on the brane which 
would be sitting in a trace free stresses induced by the Weyl curvature 
of the bulk.
\end{abstract}


\vspace{0.5cm}

 In view of the developments in the string theory, the study of gravity in 
dimensions higher than the usual four has been in vogue for quite a 
while. The recent results indicate that extra dimensions need not be 
compact, they could be large and even infinite [1,2]. In the Randall - 
Sundrum (R - S) model [2], the 4-D Universe is a 3-brane acting as a domain 
wall between the two semi infinite 5-D negative $\lambda$ regions. That is, 
the actual Universe is embedded in the 5-D bulk satisfying the appropriate 
boundary conditions. All 
non-gravitational matter fields remain localized in the brane while free 
gravitational field and gravitational waves also propagate off the brane in 
the 5-D bulk. The free field in the bulk backreacts on the brane to induce a 
trace free stress tensor through the 5-D Weyl tensor evaluated at the 
brane. The Einstein field equation for the brane has been formulated by 
Shiromizu, Maeda and Sasaki [3].

 In the absence of energy distribution on the brane, the stress 
tensor on the brane would have only the trace free Weyl term which need 
not a priori satisfy the usual positive energy conditions. The energy density 
of the induced stresses could as well be negative. The effective stress tensor
on the brane could thus legitimately include negative energy. It turns out 
that for a static black hole on the brane, the induced energy from the free 
gravitational field in the bulk must be negative to contribute positively to 
the hole's gravity (attractively). In this context we would like to argue 
that the proper energy condition should be that the induced 
energy must be negative to contribute attractively in unison with the hole. 
This should be true in general but we shall establish this explicitly for a 
static source.

 The basic criterion for energy conditions is that they should ensure 
the usual behaviour for matter which we observe in the real life 
experience directly or indirectly. For instance, gravitational field  
produced by any kind of energy distribution must be attractive is one such 
condition. This is the case for non-gravitational energy distribution. 
However it translates for the gravitational field energy that it is 
negative. The field energy is non-localizable so could also be 
non-gravitational energy distribution like the electric field energy. 
Now the question is, what norm should be applicable to such a distribution; 
positive with the non-gravitational energy or negative with the field 
energy? We would argue that the latter is the case and would define 
the proper energy condition for non-localizable energy is that it be 
negative. That is, it would contribute attractively only when it is 
negative. That is, positive energy is repulsive. 

 This has direct relevance and application for the induced energy on 
the brane from the bulk. It has zero trace and hence it cannot be localized. It must follow the norm of the field energy. It must therefore be 
negative to produce attractive gravity on the brane. This will in turn determine 
the sign of the ``tidal'' charge.  

 Let me first give a couple of known examples of positive energy producing 
repulsive effect. Both for the charged and the Schwarzschild - de Sitter 
black holes, the electric field as well as the cosmological constant produce 
positive energy density, which produces repulsive gravity. It should not 
therefore come as surprise, however such an explicit statement is perhaps 
being made for the first time. Very recently, Vollick [4] has argued that 
energy density induced by the 5-D vacuum solution onto the brane would be 
negative. Negative energy has been considered in quantum field theory for 
long time [5]. It has also been argued that it should also satisfy certain 
inequalities [6] restricting the duration of occurrence when observed. It 
turns out that more negative the density is shorter would be its life time. 
  
 Consider a simple but intuitively and physically illuminating situation. 
Let an isolated particle of mass $M$ sit in a non-gravitational energy 
distribution which falls off to zero asymptotically. That is, the total 
energy at infinity is $M$, while the energy contained inside some 
radius $R$ would be $M - E(r>R)$. In this, the energy lying exterior to 
$R$ has been subtracted out from the energy at infinity, $E(\infty) = M$. 
The Newtonian potential would then be given by,
\begin{equation} 
 {\bf\phi}(R) = - \frac{M - E(r>R)}{R}
\end{equation}
and the acceleration would be
\begin{equation}
 a = - \phi ^\prime = - \frac{M - E}{R^2} - \frac{E^{\prime}}{R}.
\end{equation}

 Since $E$ decreases with $r$, the second term would be 
positive implying repulsive gravity unless of course, $E$ itself is 
negative. Thus the energy distribution surrounding a mass point 
must have negative energy to produce attractive gravity in unison 
with the mass point. This clearly illustrates the important point that 
energy distribution engulfing the particle would act attractively in unison 
only when it is negative. We could apply this argument straightaway 
 to the Reissner - Nordstr\"om solution of a charged black hole [7]. It 
is quite illuminating to see that the electric field energy would produce 
repulsive effect.

 We shall now establish this result for a static source in general 
relativity (GR) and also formulate the proper energy condition for 
non-localized energy. In GR non-gravitational matter/energy as well as the 
gravitational field energy has gravitational charge. The difference in 
their action is that the former imparts a pull through the gradient of 
potential on the test particle while the latter warps/curves the space 
around the gravitating body [8]. However the two must work in resonance. 
That is the space warping must be in line with the pull.

 For a static source, the spacetime metric is given by
\begin{equation} 
ds^2 = - A dt^2 + B dr^2 + r^2 (d\theta^2 + sin^2\theta d\phi^2)
\end{equation}
where $A$ and $B$ are functions of the radial coordinate $r$. The 
gravitational potential sits in $A$ and imparts gravitational attraction 
on test particle which is given by $- A^{\prime}/2A = - M/r^2$. On the other 
hand the space curvature has the following non-zero Riemann components, 
\begin{equation}
R^{r\theta}_{r\theta} = R^{r\phi}_{r\phi} = \frac{B^{\prime}}{2ABr} = - 
M/r^3, \hspace{0.3cm}
R^{\theta\phi}_{\theta\phi} = \frac{1 - 1/B}{r^2} = 2M/r^3
\end{equation}
where the terms with $M$ refer to the Schwarzschild solution. 

 From the Einstein equation, we can write for the energy density,
\begin{equation}
\rho = \frac{B^{\prime}}{B^2r} + \frac{1}{r^2}(1 - \frac{1}{B})
\end{equation}

 Note that the 3-space has the topology of 
$S^2XR$. From the point of view of radial motion, we can specialize to 
the equatorial plane. The metric for the Schwarzschild case, $A = B^{-1} = 1 - 2M/r$, would  read as
\begin{equation}
(1 - 2M/r)^{-1} dr^2 + r^2 d\phi^2
\end{equation}
which has the negative curvature $-M/r^3$. It can be embedded in the 3-flat 
space by writing 
\begin{equation}
z^2 = 8Mr - 16M^2
\end{equation}
which is a parabola and would generate a paraboloid of revolution (see 
Fig.1). Clearly it has negative curvature which would tend particles to 
roll down (or oppose motion leading to increasing $r$) to decreasing value 
of $r$ in unison with the attractive action of the potential. That is 
negative curvature here is the analogue of the attraction in the equation 
of motion.

 Thus the norm for the curvature is set that it should be negative. The 
sign of the relevant curvature is determined by $B^{\prime}$ which must 
be negative. From eq.(5), it is  clear that negative density 
would contribute negatively to the curvature and consequently 
positively (in line with the potential) to the field. The positive 
density would contribute positively to the curvature, opposing the 
potential, thereby weakening the field though overall curvature may still 
be negative as is the case for the charged black hole.

 Corresponding to the positive energy conditions for the matter fields, 
the ``positive'' energy condition for the non-localized energy distribution 
in line with the gravitational field energy is that it be negative. We 
have used the prototype Schwarzschild field to ellucidate the norm set by 
gravitational field energy. The positivity of energy could be 
equivalently be stated as negativity of the field energy which is further 
equivalent to negativity of space curvature in ($r,\theta$) plane. 

 This alternative formulation of the energy condition is particularly 
pertinent to the R - S brane world scenario [2]. Here, the 
free gravitational filed reflected from the bulk onto the brane through 
the Weyl curvature of the bulk would give rise to trace free stress 
tensor on the brane. It would represent a non-localizable energy 
distribution and hence the only energy condition which is pertinent in 
this case is the one we have formulated above. That is, it be 
negative to warp/curve the space in harmony with the action of the 
gravitational potential. 
 
 This has direct relevance and application for black hole on the brane, 
which would always have some non-zero energy distribution around it coming 
from the bulk Weyl curvature. It is therefore important that the induced energy 
density must be negative if it were to contribute positively (attractively) 
to the hole's field. This should really be the case because only the 
free field propagates in the bulk which would have negative energy and 
consequently it would reflect back negative energy density on the brane. 
Since the induced stresses are trace free, the energy distribution is 
non-localizable and hence it should follow the negativity norm of the 
gravitational field energy. Recently, a solution for a black hole on the 
brane has been proposed [9] which is described by the 4-D 
Reissner-Nordstr\"om metric for the charged black hole. Of course there 
is no electric charge, the corresponding parameter refers to the "tidal" charge 
induced by the bulk Weyl curvature. This charge parameter would, in 
contrast to the electric charge case, be negative because the gravitational 
field energy in the bulk is negative and to satisfy the proper 
{\it negative energy condition}. This solution would be valid only near 
the horizon. The high energy corrections to the Schwarzschild solution 
should only introduce modification to the field without altering its basic 
character. The "tidal" charge would therefore have to be negative. If 
it is not so, it would give rise to both the event as well as the Cauchy 
horizon which will change the singularity character of the static black hole. 
What is expected is a modification retaining the basic character of the 
spacetime. Thus the tidal charge parameter and the energy of the distribution 
immersing the hole must be negative. 

 All this discussion of negativity of energy  would have direct 
application to the study of gravitational collapse on the brane. As the 
high energy corrections to GR according to the R-S model [2] would 
manifest in the induced traceless  
energy distribution on the brane which would be negative. It would therefore 
act in unison with the 
gravitational attraction and would strengthen the collapse. They  
would act in line with the cosmic 
censorship hypothesis for collapse on the brane. It would be interesting 
to investigate the cases on the brane that lead to a naked singularity in GR 
[10]. Would the singularity still persist or not? The answer to this 
question would be of significance for the ultimate outcome of
gravitational collapse. 

{Acknowledgements:} This work was started at the Albert Einstein 
Institute, Golm and carried through to the universities of Wales, 
Cardiff, of the Basque Country, Bilbao and of Portsmouth, and Queen Mary 
and Westfield College, London. I wish to thank all these institutions and 
my hosts for warm hospitality. I would also like to thank T. Shiromizu, Roy 
Maartens and Sanjeev Dhurandhar for useful discussions.

\begin{figure}
\centering
{\epsfysize=8.5cm
{\epsfbox[0 0 500 250]{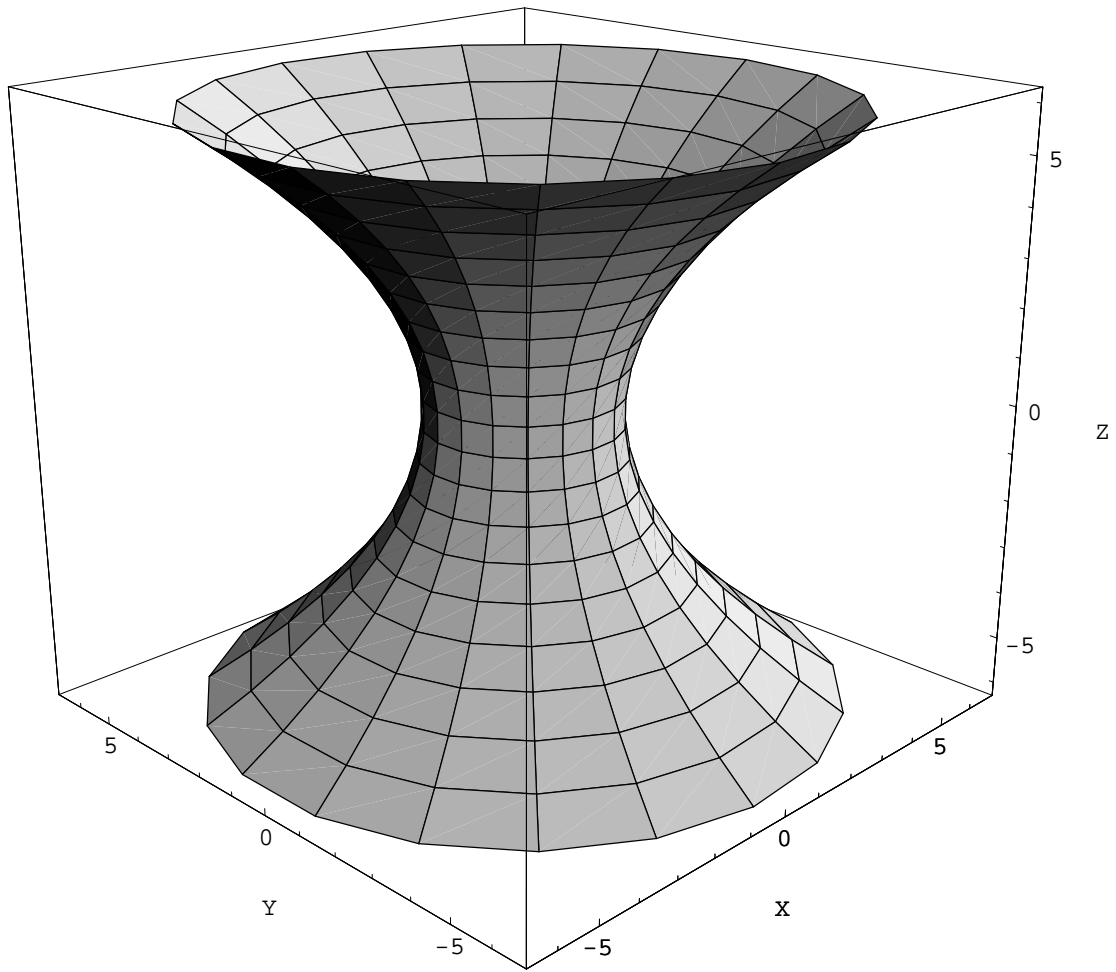}}} 
\caption[]{It shows the paraboloid of revolution for the parabola
 $z^2 = 8 r - 16$, where $r^2 = x^2 + y^2$.}
\end{figure}

\end{document}